\documentclass[letterpaper, 10 pt, conference]{ieeeconf}  

\IEEEoverridecommandlockouts                              
\usepackage{graphicx} 

\title{\LARGE \bf RapidLearn: A General Purpose Toolkit for Autonomic Networking}

\author{\textbf{Jatin Sharma$^*$, Nikhilesh Behera$^*$, Priya Venkatraman$^*$, Boon Thau Loo}\\ 
School of Engineering \& Applied Science\\
        University of Pennsylvania\\
        \{jatin, nbehera, vpriya, boonloo\}@seas.upenn.edu\\
        \hspace*{ 0.5 in}
}

\usepackage{listings}
\usepackage{color}

\definecolor{dkgreen}{rgb}{0,0.6,0}
\definecolor{gray}{rgb}{0.5,0.5,0.5}
\definecolor{mauve}{rgb}{0.58,0,0.82}

\providecommand{\keywords}[0]{\textbf{\textit{Index terms---}}}
\begin{document}

\maketitle
\thispagestyle{empty}
\pagestyle{empty}

\def\thefootnote{*}\footnotetext{These authors contributed equally to this work}

\begin{abstract}

Software Defined Networking has unfolded a new area of opportunity in distributed networking and intelligent networks. There has been a great interest in performing machine learning in distributed setting, exploiting the abstraction of SDN which makes it easier to write complex ML queries on standard control plane. However, most of the research has been made towards specialized problems (security, performance improvement, middlebox management etc) and not towards a generic framework. Also, existing tools and software require specialized knowledge of the algorithm/network to operate or monitor these systems. We built a generic toolkit which abstracts out the underlying structure, algorithms and other intricacies and gives an intuitive way for a common user to create and deploy distributed machine learning network applications. Decisions are made at local level by the switches and communicated to other switches to improve upon these decisions. Finally, a global decision is taken by controller based on another algorithm (in our case voting). We demonstrate efficacy of the framework through a simple DDoS detection algorithm.

\end{abstract}

\section{INTRODUCTION}

Software Defined Networking has been proven to be a great tool for network management due to (i) decoupling of network hardware from software, (ii) centralized control over switches and middleboxes, (iii) network function virtualization, and (iv) traffic engineering. SDN has reduced the operating costs incurred by service providers while providing better control over traditional services and flexibility to add new services.
 
Recently, there has been increased interest in academic and research communities in developing intelligent networks using distributed machine learning over the simple network elements. The idea is similar to Active Networks in concept but takes an SDN approach. 


Most of the work so far has been problem specific (security, performance improvement, middlebox management etc) and no generic framework exists that can be used for all of them simultaneously. These tools require domain specific knowledge to be operated and analysed. In this paper, we present an idea to perform Machine Learning in distributed setting through a simple declarative programming model. We extend RapidNet, an existing rapid networking protocol simulator, with distributed machine learning capabilities so that network researchers can write complex Machine Learning queries over a network in a decoupled fashion through the control plane abstraction. We also show how local and global decisions can be made at different levels of network. These local and global decision algorithms can, further, be customized by the user as per his use giving him full control over the network. The toolkit abstracts out the underlying structure, algorithms and other intricacies providing an intuitive platform for a user to create and deploy distributed Machine Learning network applications. Decisions made at local level by the switches can be communicated to other switches to reinforce these beliefs and, to finally, reach a global decision.We demonstrate efficacy of the framework through a simple DDoS detection algorithm working on a two SDN networks topology.

Our aim is to present a proof-of-concept framework and demonstrate effectiveness of the toolkit in reducing effort, time and enhancing performance, scale. Benchmarking accuracy, precision and recall of machine learning algorithms is not the aim. These algorithms are declared by the end user as per his requirements.

In later sections, we browse through some of the existing work in this area, discuss methodology used for the framework, individual project components and design decisions and some tests and evaluation results. 


\section{PROJECT COMPONENTS}

\begin{figure}
      \centering
      \includegraphics[width=70mm]{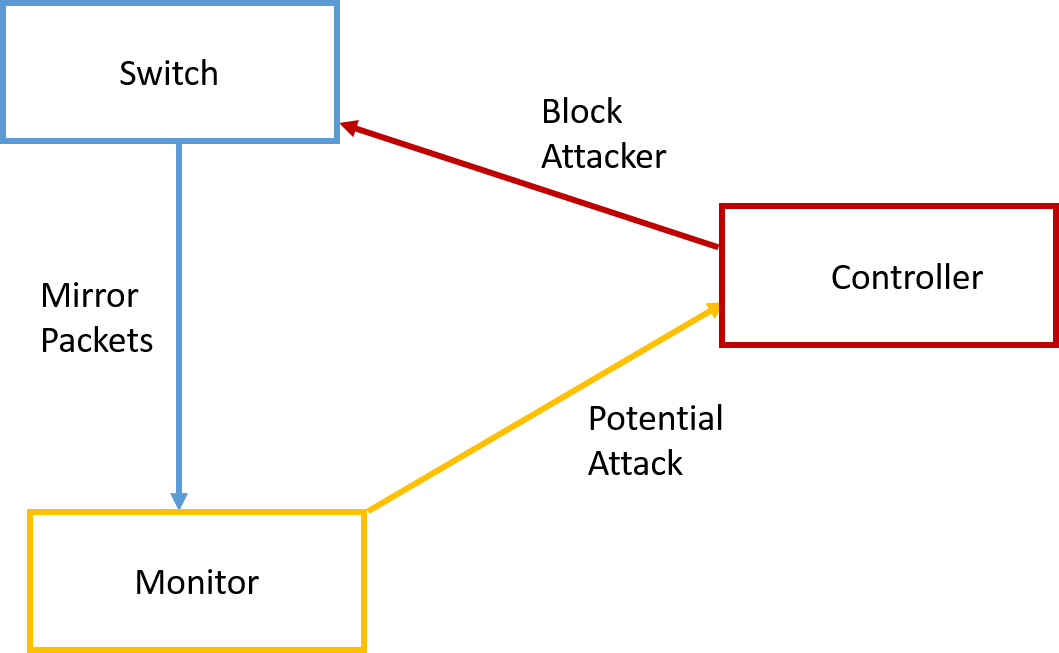}
      \caption{Message flow among RapidLearn components}
   \end{figure}
   
We defined end-to-end network topology in RapidNet using declarative NDLog program. A typical NDLog program specifies the rules to be executed at a network component and messages that needs to be passed to other components. We enable network switches to perform machine learning through a low-latency neighbor element called "Monitor". A monitor is a typical top of the rack VM that can execute any program. These monitors communicate within themselves and with the centralized controller to perform Distributed and hierarchical learning among switches and controller.

\subsection{Host}
A host is an end user machine connected to the network. In our usecase, a host would typically be a server, a legitimate client user or a malicious attacker. A host can send packets to the network and it can also receive packets from the network. 

\begin{lstlisting}
/*Host program*/
/*Packet initialization*/
rh1 packet(@Switch, Host, SrcMac, DstMac) :-
 initPacket(@Host, Switch, SrcMac, DstMac),
 link(@Host, Switch, OutPort).

/*Receive a packet*/
rh2 recvPacket(@Host, SrcMac, DstMac) :-
 packet(@Host, Switch, SrcMac, DstMac),
 link(@Host, Switch, InPort).
\end{lstlisting}

\begin{figure}
      \centering
      \includegraphics[width=70mm]{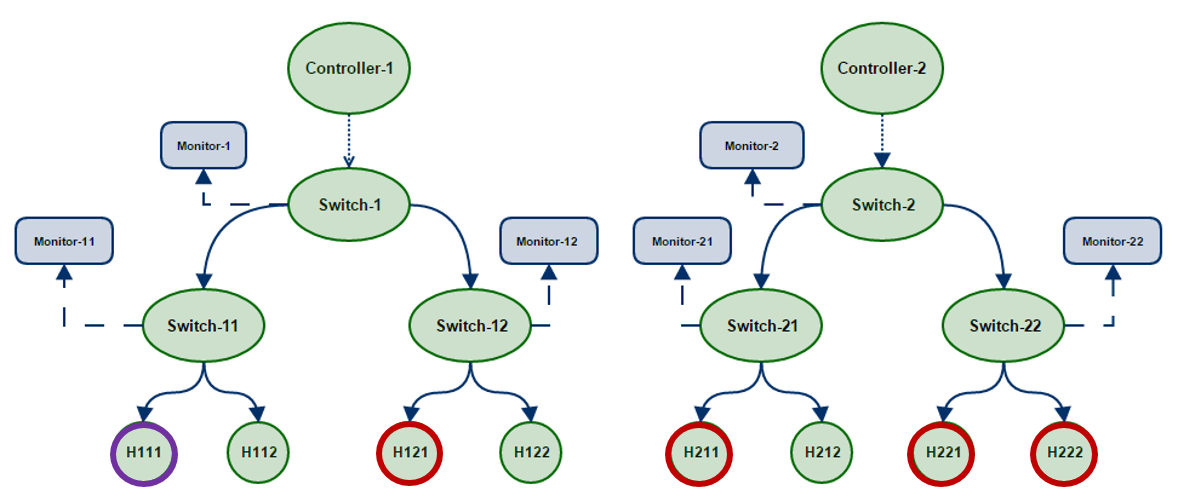}
      \caption{Type of hosts (i) Attackers(red) and (ii) victim (purple)}
   \end{figure}

\subsection{Switch}

A switch is the basic component for our network topology. It has a MAC-learning capability. It is connected to an OpenFlow controller. When a switch receives a packet it looks into its forwarding table for next hop. On a hit, it forwards the packet on the appropriate port. However, on a miss, it forwards the packet to the controller and waits for the controller to make a decision and set appropriate entry into switch's forwarding table. We add one more component in this simple flow - a monitor. A monitor is a top-of-the-rack VM that connects to the switch with very low latency, and can handle compute heavy tasks. The switch mirrors all its incoming traffic to the associated monitor.

\begin{lstlisting}
/*Switch program*/
/*Query the controller when receiving unknown packets */
rs1 matchingPacket(@Switch, SrcMac, DstMac, InPort, TopPriority) :-
 packet(@Switch, Nei, SrcMac, DstMac),
 link(@Switch, Nei, InPort),
 maxPriority(@Switch, TopPriority),
 blockedLinks(@Switch, Nei, Port),
 Port != InPort.

/*Recursively matching flow entries*/
rs2 matchingPacket(@Switch, SrcMac, DstMac, InPort, NextPriority) :-
 matchingPacket(@Switch, SrcMac, DstMac, InPort, Priority),
 flowEntry(@Switch, MacAdd, OutPort, Priority),
 Priority > 0,
 DstMac != MacAdd,
 NextPriority := Priority - 1.

/*A hit in flow table, forward the packet accordingly*/
rs3 packet(@OutNei, Switch, SrcMac, DstMac) :-
 matchingPacket(@Switch, SrcMac, DstMac, InPort, Priority),
 flowEntry(@Switch, MacAdd, OutPort, Priority),
 link(@Switch, OutNei, OutPort),
 Priority > 0,
 DstMac == MacAdd.

/*If no flow matches, send the packet to the controller*/
rs4 ofPacket(@Controller, Switch, InPort, SrcMac, DstMac) :-
 ofconn(@Switch, Controller),
 matchingPacket(@Switch, SrcMac, DstMac, InPort, Priority),
 Priority == 0.

/*Insert a flow entry into forwarding table*/
rs5 flowEntry(@Switch, DstMac, OutPort, Priority) :-
 flowMod(@Switch, DstMac, OutPort),
 ofconn(@Switch, Controller),
 maxPriority(@Switch, TopPriority),
 Priority := TopPriority + 1.

/* Set Max Priority */
rs6 maxPriority(@Switch, Priority) :-
 flowEntry(@Switch, DstMac, OutPort, Priority).

/*Following the controller's instruction, send out the packet as broadcast*/
rs7 packet(@OutNei, Switch, SrcMac, DstMac) :-
 broadcast(@Switch, InPort, SrcMac, DstMac),
 link(@Switch, OutNei, OutPort),
        OutPort != InPort.

/*Send every incoming packet to the monitor*/
rs8 monPacket(@Monitor, Controller, Switch, SrcMac, DstMac) :-
        monconn(@Switch, Monitor),
        ofconn(@Switch, Controller),
        packet(@Switch, Nei, SrcMac, DstMac).

/* If Controller sends block then call blockHost */
rs9 blockedLinks(@Switch, Nei, InPort) :-
        block(@Switch, Nei, SrcMac, DstMac),
        link(@Switch, Nei, InPort),
        ofconn(@Switch, Controller).
\end{lstlisting}

   \begin{figure}
      \centering
      \includegraphics[width=70mm]{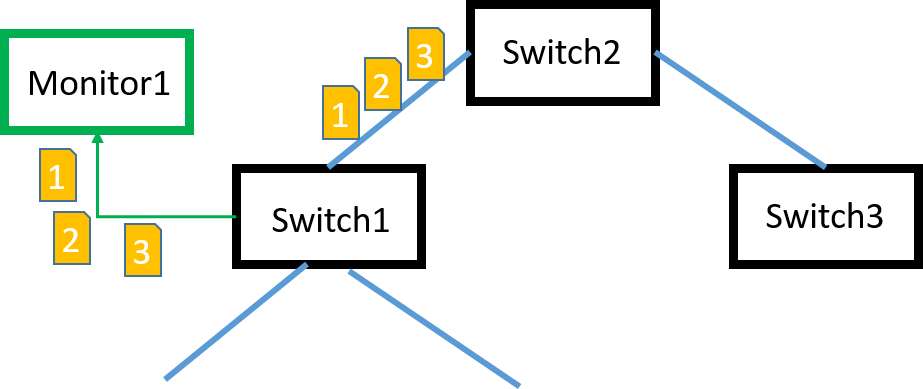}
      \caption{A typical switch in RapidLearn}
   \end{figure}

\subsection{Monitor}
As mentioned above, the monitor connects to the switch with very low latency, and can handle compute heavy tasks. Monitor receives a continuous stream of packets from its associated switch. The monitor splits the packets into session streams from different users so that an attacker can be differentiated from a legitimate user. It then apply Machine Learning model to detect on each ongoing session to detect any anomaly. We use Support Vector Classifier (C-SVC) at the monitor for anomaly detection. Once SVC detects an ongoing attack from one or more of the users the monitor can take two actions - (i) directly send its belief to the controller, or (ii) message other monitors to enhance its belief. All Monitors send their local decisions to the controller. Independent of these two actions a monitor can also choose to rate-limit the link with attackers traffic until a global decision is made by the controller.

\begin{lstlisting}
/*Monitor program*/
/*If attack is happening. Tell the controller about Switch, SrcMac and DstMac*/
/*update this to send the status to other monitors*/
rm1 ddos_yes(@Controller, Switch, SrcMac, DstMac) :-
        monPacket(@Monitor, Controller, Switch, SrcMac, DstMac).
        ddos_check(Switch, SrcMac, DstMac) == 1.
\end{lstlisting}

   \begin{figure}
      \centering
      \includegraphics[width=70mm]{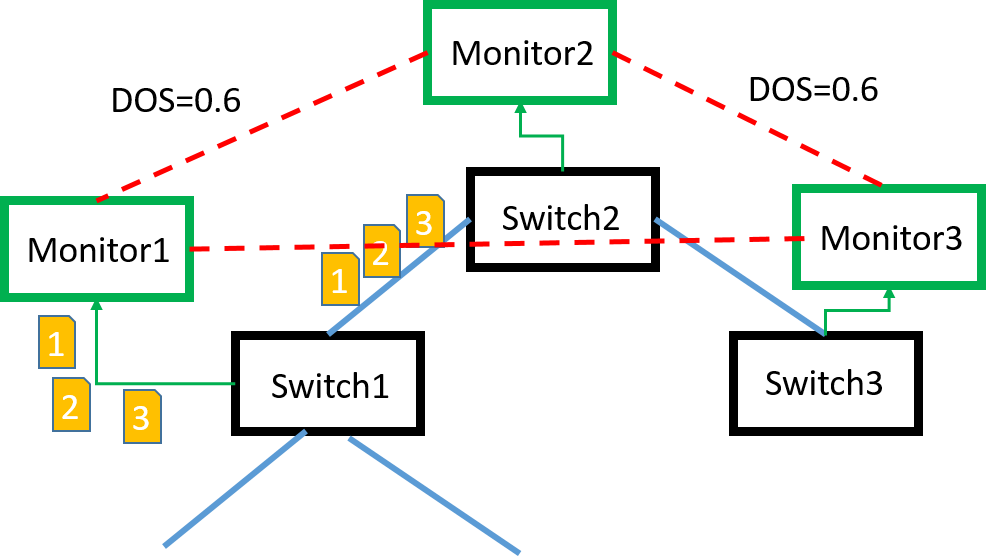}
      \caption{Multi-monitor collaboration for iterative decisions}
   \end{figure}

\subsection{Controller}

We use a simple Openflow controller that makes forwarding decisions and modifies switch's forwarding table. Apart from this general behavior, switch also listens from several monitors about possible ongoing attacks and make global decisions to block an attacker. We emulated the Controller capability in RapidNet through a sequence of rules.

\begin{lstlisting}
/*Controller program*/
/*Install rules on switch*/
rc1 flowMod(@Switch, SrcMac, InPort) :-
 ofconn(@Controller, Switch),
 ofPacket(@Controller, Switch, InPort, SrcMac, DstMac).

/*Instruct the switch to send out the unmatching packet*/
rc2 broadcast(@Switch, InPort, SrcMac, DstMac) :-
 ofconn(@Controller, Switch),
 ofPacket(@Controller, Switch, InPort, SrcMac, DstMac).

/*If monitor indicates an ongoing attack then block it*/
rc3 block(@Switch, SrcMac, DstMac) :-
        ddos_yes(@Controller, Switch, SrcMac, DstMac).

\end{lstlisting}

   \begin{figure}
      \centering
      \includegraphics[width=70mm]{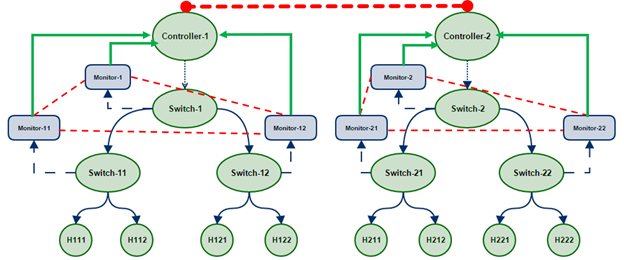}
      \caption{A simple two network topology for DDOS detection and mitigation}
   \end{figure}


\section{METHODOLOGY}
   
We used RapidNet to emulate a simple 2-network SDN Switch-Controller topology as shown in fig1. Each network consists of multiple end hosts connected to SDN enabled switches. These switches further connect to more switches in a hierarchical arrangement. It is to emulate a typical multi-level switch arrangement in real networks. 

We assume that each switch has MAC-learning capability. Every switch with a particular network (administrative domain) has a known controller to which it sends packets at the beginning of the flow to make flow decisions. These controllers will eventually obey openFlow messages to and fro the switches. Controllers make high level decisions whether to block a flow, allow it on a particular port or broadcast it.

We then introduce, Monitors as an active component of the network that would do most of the heavy-lifting by executing Machine Learning classifier to detect potential DDOS attacks. Monitors are expected to be collocated with the switch resulting into a very low-latency communication. This allows monitor to run a VM that can run any kind of machine learning algorithm. In our illustrative use-case, we run Support Vector Classifier (C-SVC) on the monitor but it can be customized to run any learning algorithm suitable for the problem at hand. This provides a great flexibility to the protocol developer or researcher. 

These monitors make decisions in two disjoint phases. In first phase, they keep monitoring the mirrored packet stream from the switch. This stream is partitioned into multiple slices. Every slice corresponds to an active but distinct user session. Slicing the packets stream is required to separate legitimate users from malicious users. Whenever an anomalous activity is monitored from one of the hosts that host can be blocked without affecting all other hosts communicating with the same switch. This phase is called Local Decision.

In second phase, the monitors can either directly send their beliefs to the controller or iteratively enhance their beliefs by collaborating with each other. The controller, finally, receives beliefs from all the monitors and makes final decision. Controller can use ensemble learning or just deploy a simple voting mechanism at the end. We followed latter approach where each monitor sends its belief to the controller and controller blocks a host based on majority decision. This phase is called Global Decision. 

\begin{figure}
      \centering
      \includegraphics[width=70mm]{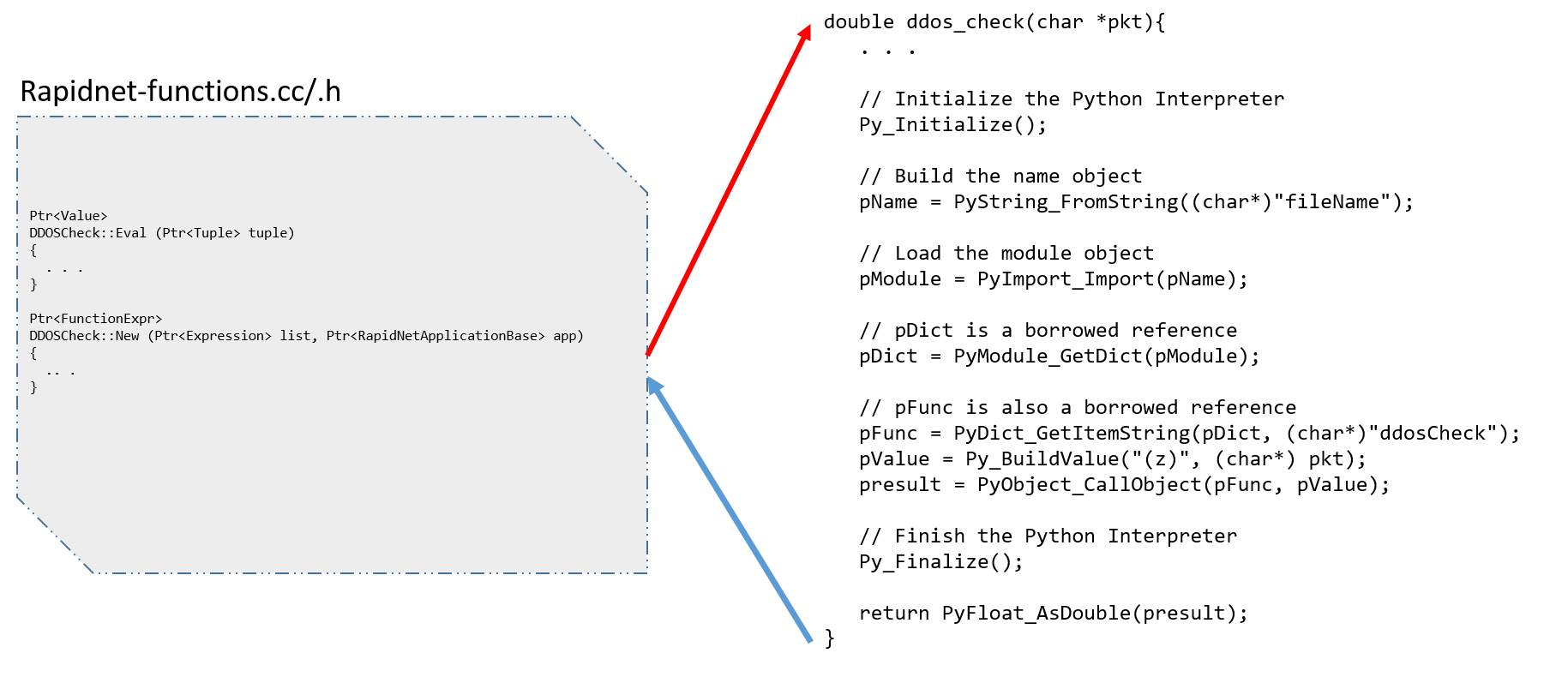}
      \caption{RapidLearn calls python script to execute SVC}
   \end{figure}
   
We wrote a python script that runs C-SVC over the packet stream on a monitor. This script is called by a user defined function in rapidNet.


\section{Autonomic Learning}

\subsection{Dataset}
We used Caida dataset to train our learning algorithm. Caida dataset contains approximately one hour of anonymized traffic traces that include the attack traffic on the victims and the response from the victims on August 4, 2007. The total size of the original dataset is 21 gigabytes. The type of the file is the packet capture file. Since there are few applications to read these data, we exported all the data to comma-separated value files for easy data reading in any programming language and machine. 

The size of the dataset that we obtained is big and analyzing all data takes very long time. So, to analyze the dataset efficiently and quickly, we extracted some features including source IP address, destination IP address, time interval in seconds between packets, and packet size in bytes from the dataset. We analyzed the total number of packets that were sent from the attacker and victim, the mean time intervals in seconds, and mean packet size in bytes for each packet. 

\begin{figure}
      \centering
      \includegraphics[width=70mm]{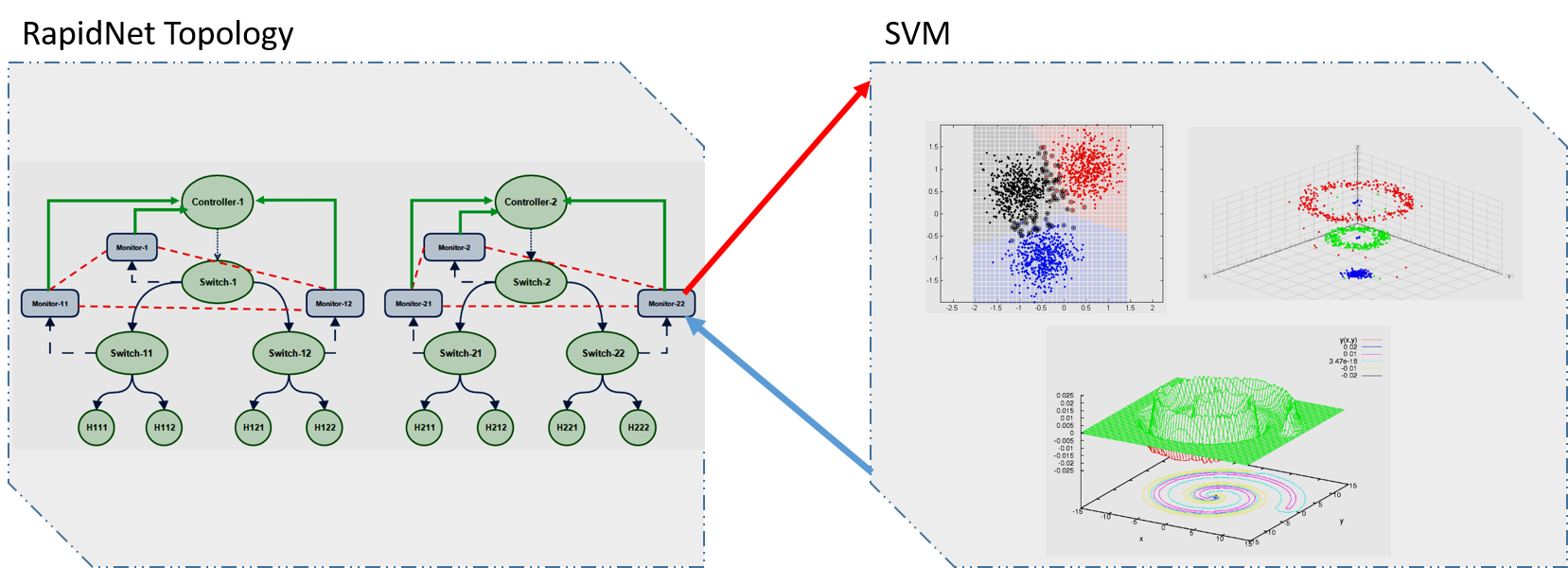}
      \caption{Machine Learning flow in RapidLearn}
   \end{figure}
   
\subsection{Training}
We trained and tested three types of dataset using C-SVC with the RBF (Gaussian) kernel. The precision for each dataset is 0.98. These numbers mean that the detection system could successfully predict the attack. These results show that the detection accuracy is high when the system should be alerted. However, the recall is 0.99. These results show that the system predicted some normal data as the attack data.

\section{EVALUATION}

\subsection{Testing}
\subsubsection{Validation of the network containing switches and end hosts}
The designed topology was tested for connectivity between switches and end hosts via ping packets being transferred from one end host to another.

\subsubsection{Valid packet/data capture for the training dataset}
DARPA dataset was used for training and testing purposes.

\subsubsection{Testing as to whether the rules are being installed correctly via NDLog}
NDLog rules were applied on switches and were tested accordingly for installation and uninstallations of rules on switches

\subsubsection{Reliable message passing between switches and controller}
Message passing between switches were made reliable and idempotent.

\section{CONCLUSIONS}

In this paper, we proposed one approach using machine learning to develop an DDoS attack detection system. We analyzed a large number of network communication packets, and implemented a DDoS attack detection system using the patterns of DDoS attacks for each IP address. We presented and discussed the difficulties involved in developing a DDoS attack detection system using packets and features of DDoS attacks that are important to detect an attack. These features are very helpful in developing a detection system for DDoS attacks. Analyzing the packets, we calculated bytes per second with the time sequence and found that there are mainly two periods in DDoS attack.We also demonstrated a POC framework which leverages Software Defined Networking paradigm to execute Distributed Machine Learning algorithm over a real network. User doesn't need any domain specific knowledge can can quickly create test topologies and deploy his experimental protocols. He can utilise beliefs at various levels to finally reach global decisions. We demonstrated an example use-case of two network DDoS detection protocol and showed how it hugely reduces time, effort and specialized knowledge while improving scale, performance and reliability.

\addtolength{\textheight}{-12cm}   

\section{ACKNOWLEDGEMENT}

We thank Anne, Anduo and Chen for their continuous support and comments that greatly improved the manuscript.


\end{document}